\documentstyle[12pt]{article}
\newcommand{\newc}{\newcommand}    
\newc{\ra}{\rightarrow} 
\newc{\lra}{\leftrightarrow} 
\newc{\beq}{\begin{equation}} 
\newc{\eeq}{\end{equation}} 
\newc{\barr}{\begin{eqnarray}} 
\newc{\earr}{\end{eqnarray}} 
\newc{\texa}{\textstyle}
\newc{\paral}{\parallel}
\newc{\und}{\underline}
\newc{\pars}{\partial}
\newc{\nonu}{\nonumber \\}
\newc{\jump}{\nonumber \\[2.0mm]} 
\newc{\rar}{\rightarrow}
\newc{\al}{\alpha}

\begin{document}
\thispagestyle{empty}
\begin{center}
{\Large \bf Axisymmetric equilibria of a gravitating plasma \vspace{2mm} \\
 with incompressible flows
	    \vspace{5mm} }\\
\large\bf
{\large  G. N. Throumoulopoulos
\footnote{Permanent  address: 
University of Ioannina, Association Euratom - Hellenic Republic, 
Physics Department, Theory Division, GR  451 10 Ioannina, Greece} 
and H. Tasso             
\vspace{1mm}\\
{\it   Max-Planck-Institut f\"{u}r Plasmaphysik, EURATOM
Association \\ \vspace{1mm}
 D-85748 Garching, Germany }
\\ \vspace{2mm}
November  2000}  
\end{center}
\vspace{1mm}
%
\begin{center}
{\large\bf Abstract} 
\end{center}

It is found that the ideal magnetohydrodynamic
equilibrium of an  axisymmetric  gravitating
magnetically confined plasma with incompressible flows is governed by 
a second-order elliptic
differential equation for the poloidal magnetic flux function
 containing five flux functions
coupled with a Poisson equation for the gravitation potential,
and an algebraic relation for the pressure. This set of equations   
is amenable to analytic solutions.
As an application, the 
 magnetic-dipole static axisymmetric equilibria with vanishing
 poloidal plasma currents derived recently by 
 Krasheninnikov,
Catto, and Hazeltine [Phys. Rev. Lett. {\bf 82 }, 2689 (1999)] are extended
to plasmas with finite poloidal currents,
subject to 
gravitating  forces from  a massive body (a star or 
 black hole) and 
inertial forces due to incompressible sheared flows. Explicit solutions 
are obtained
in two regimes:  (a) in the low-energy regime
$\beta_0\approx \gamma_0\approx \delta_0 \approx\epsilon_0\ll 1$, where  
$\beta_0$, $\gamma_0$,
$\delta_0$, and $\epsilon_0$ 
are related to the  
 thermal, 
 poloidal-current,
flow and gravitating energies normalized to the 
poloidal-magnetic-field energy, respectively,
and (b) in the high-energy regime 
$\beta_0\approx \gamma_0\approx \delta_0 \approx\epsilon_0\gg 1$. 
It turns out that 
in the high-energy
regime all four forces, 
 pressure-gradient,  toroidal-magnetic-field, 
 inertial, and gravitating contribute equally to the
 formation of  magnetic surfaces  
very 
extended and 
localized about the symmetry plane  such that the
resulting equilibria resemble the accretion 
disks in astrophysics.
\newline\newline

%
\newpage
\setcounter{page}{1}

\begin{center} 
{\large \bf 1.\ \ Introduction}
\end{center}            
The difficult problem of equilibrium with flow  
has been  the
subject of  an increasing number of investigations (e.g., 
in
relation to the present work see Refs. \cite{MoSo}-\cite{Bu}, \cite{KrCaa},
\cite{ThTa99}) 
on both  laboratory and astrophysical 
plasmas.    
Even for vanishing gravity 
the ideal magnetohydrodynamic (MHD) equilibrium of a
symmetric (two dimensional) plasma with arbitrary compressible
flows associated, e.g.  with isothermal magnetic surfaces, 
  satisfies a second-order
partial differential equation for the poloidal magnetic flux function
$\psi$
 coupled  with an
algebraic Bernoulli  equation for the density \cite{MoSo}. 
Depending on the value of the 
poloidal Mach number $M^2$ (defined in Sec. 2), the above mentioned 
 equation can be either elliptic or hyperbolic.  
Experimental
evidence \cite{BrBi,ScBi,MaRe,Bu}, however,  confirms that (a) both density 
and temperature are 
to a very good approximation flux functions, i.e. functions 
constant
on magnetic surfaces, and (b) 
the poloidal flows (e.g. involved in the transition from the low
to the high confinement regime in tokamaks) 
lie within
the first elliptic region.   
Similar conditions may also prevail in astrophysical plasmas at least at 
distances not close 
to centers of gravity.
For magnetically 
confined plasmas the equation is expected elliptic because  
hyperbolicity, associated
with shock waves, would imply open magnetic surfaces and thereby 
abrupt confinement degradation. 
In this respect,   equilibria with
incompressible flows for which the differential equation 
becomes  always  elliptic has been of particular interest    
for both laboratory \cite{BaCh,AvBh,AnChe,ThTa,TaTh} 
and astrophysical \cite{Ts,ViTs,ViFe,PeNe} plasmas.

In previous studies \cite{ThTa,TaTh} we 
considered ideal MHD equilibria with incompressible flows
in cylindrical and toroidal geometries.
For a toroidal plasma we found that the  differential equation
 decouples from the pressure relation, thus making 
the problem analytically solvable. 
Several classes of analytic solutions
of  linearized versions of the above mentioned
differential equation associated with sheared flows were also obtained. 
For vanishing flows this equation reduces to the
Grad-Schl\"uter-Shafranov equation.
Analytic solutions of the {\em nonlinear} Grad-Schl\"uter-Shafranov
equation for a plasma with vanishing poloidal current 
at either low or high pressure confined by a dipolar  
magnetic field were obtained recently in Ref.  
\cite{KrCaHa}. 
These studies were then extended to equilibria with
purely toroidal flow \cite{KrCaa}
(being inherently incompressible because of axisymmetry),
to  gravitating  magnetic dipolar
plasmas without flow \cite{KrCab},  
and to plasmas with anisotropic pressure \cite{KrCa00}. 

The purpose of the present work is twofold: (a) to extend our 
equilibrium equations \cite{TaTh} to the case of a gravitating plasma 
with incompressible flows,  
and (b) with employment of the separable eigenvalue technique introduced
in Refs \cite{KrCaHa}-\cite{KrCa00} to derive analytic magnetic dipolar equilibria 
for a plasma at finite pressure and  poloidal current  
with  incompressible sheared flows having 
 non-vanishing 
toroidal and poloidal components,  under the exertion 
of gravitational forces
from a massive body (a star or a black hole).

In Section 2 the equilibrium equations for an axisymmetric gravitating
magnetically confined  plasma with incompressible 
flows are derived.
These equations are then reduced further  for a plasma confined
by the magnetic field of a point dipole in Section 3.
Analytic magnetic dipolar solutions are 
constructed in Sections 4 and 5 in the following  
regimes:  
(a) in the low-energy regime
$\beta_0\approx \gamma_0\approx \delta_0 \approx\epsilon_0\ll 1$, 
where  $\beta_0$, $\gamma$, $\delta_0$, and 
$\epsilon_0$ 
are related to the  
thermal, poloidal-current,  flow, and gravitating energies normalized 
to the  poloidal-magnetic-field energy, 
respectively,
and (b) in the high-energy regime 
$\beta_0\approx \gamma_0\approx \delta_0 \approx\epsilon_0\gg 1$. 
Finally, the conclusions are summarized in Sec. 6.

\begin{center}
{\large\bf 2.\ \ Stationary equilibrium equations for a gravitating
		  plasma} 
\end{center}

The ideal MHD equilibrium states of a gravitating  plasma
with flow  are governed by the following set of
equations, written in convenient units:
\begin{equation}
{\bf\nabla} \cdot (\rho {\bf v}) = 0 
					    \label{1}
\end{equation}
\begin{equation}
\rho ({\bf v} \cdot {\bf\nabla})  {\bf v} = {\bf j}
\times {\bf B} - {\bf\nabla} P   -\rho \nabla \Omega
					    \label{2}
\end{equation}
\begin{equation}
\nabla^2 \Omega= G_0 \rho_t
					    \label{2a}
\end{equation}
\begin{equation}
{\bf\nabla} \times  {\bf E} = 0 
					    \label{3}
\end{equation}
\begin{equation}
{\bf\nabla}\times {\bf B} = {\bf j }
					    \label{4}
\end{equation}
\begin{equation}
{\bf\nabla} \cdot {\bf B} = 0 
					    \label{5}
\end{equation}
\begin{equation}
{\bf E} +{\bf v} \times {\bf B} = 0.
					    \label{6}
\end{equation}
Here, $\Omega$, $4\pi G_0$, and $\rho_t$ are, respectively, 
the gravitation potential, the constant of gravity and 
the total density including contributions from the plasma
itself and from external mass sources. Standard notations are
used in the rest of Eqs. (\ref{1})-(\ref{6}).

For an axially symmetric plasma the divergence-free fields
can be expressed in terms of the functions 
$I(R,z)$, $F(R,z)$ and $\Theta(R,z)$ as
\begin{equation} 
{\bf B} = I \nabla \phi +
 \nabla\phi \times \nabla\psi,
					     \label{7}
 \end{equation}
\begin{equation}
{\bf j} = \Delta^\star\psi \nabla\phi - \nabla\phi \times \nabla I
					     \label{8}
\end{equation}
and
\begin{equation}
\rho {\bf v} =\Theta \nabla\phi +  \nabla\phi \times
{\nabla} F.
					       \label{9}
\end{equation}
Here, $R, \phi, z$ are cylindrical coordinates with $z$ 
corresponding to the axis of symmetry, constant $\psi$ 
surfaces are the magnetic surfaces,
and $\Delta^\star 
  \equiv R^2\nabla\cdot(\nabla /R^2)$.

Eqs. (\ref{1})-(\ref{6}) can be reduced  by means of 
certain integrals of the system, which are shown to be flux functions.
To identify three of these
quantities,  the time independent  electric field
is expressed by ${\bf E} = - {\bf \nabla} \Phi$ and the Ohm's law
(\ref{6}) is projected along $\nabla\phi$,  $\bf B$ and $\nabla \psi$, 
respectively,
yielding $F=F(\psi)$, $\Phi=\Phi(\psi)$,  and 
\begin{equation}
\frac{1}{\rho R^2}(IF^\prime-\Theta)= \Phi^\prime.
					      \label{10}
\end{equation}
A fourth flux function is derived from the component
of the force-balance equation (\ref{2})  along $\nabla \phi$:
\begin{equation}
I\left(1-\frac{(F^\prime)^2}{\rho}\right)
+R^2 F^\prime \Phi^\prime \equiv  X(\psi),
					      \label{11}
\end{equation}
the flux function $X(\psi)$  being related to the toroidal
magnetic field (see also Eq. (21) in Sec. 3). Note that
the toroidal quantities $I(\psi, R)$ and $\Theta(\psi, R)$
are not flux functions.
With the aid of the flux functions identified  
the components 
of Eq. (\ref{2}) along $\bf B$ and perpendicular to
a magnetic surface are put in the respective forms
\begin{equation}
{\bf B} \cdot \left\lbrack{\bf \nabla} \left(\frac{v^2}{2}
+ \frac{\Theta}{\rho}\Phi^\prime\right)
+ \frac{\nabla P}{\rho} + \nabla \Omega \right\rbrack = 0 
					       \label{14}
 \end{equation}
and
\begin{eqnarray} 
\left\{ {\bf \nabla} \cdot 
\left\lbrack\left(1- \frac{(F^\prime)^2}{\rho}\right)
\frac{{\bf \nabla}\psi}{R^2} \right\rbrack 
+ \frac{F^{\prime\prime}F^\prime }{\rho}\frac{|\nabla\psi|^2}{R^2}\right\}
   |\nabla \psi|^2
& & \jump 
+ \left\lbrack\frac{\rho}{2}\left(\nabla v^2
		- \frac{\nabla (\Theta/ \rho)^2}{R^2}\right)
   +  \frac{\nabla(I^2)}{2 R^2} +\nabla P  
   +  \rho\nabla\Omega \right\rbrack\cdot \nabla\psi = 0 & &
						\label{15} 
\end{eqnarray}
It is pointed out that Eqs. (\ref{14}) and (\ref{15}) hold 
for any equation of state.


We now consider incompressible flows, 
$\nabla\cdot {\bf v} = 0$, which  
 on account of Eq. (\ref{1}) implies that the density
is a flux function.   With the aid of $\rho=\rho(\psi)$,
Eq. (\ref{14}) can be integrated to yield an expression for 
the pressure, i.e.
\begin{equation} 
P = P_s(\psi) - \rho \left(\frac{v^2}{2} + \Omega
	       - \frac{R^2 (\Phi^\prime)^2}{1-(F^\prime)^2/\rho} \right),
						\label{19}    
\end{equation}
where $P_s(\psi)$ is a flux function which
for vanishing flow and gravity 
(${\bf v}=\Omega=0$) represents the static pressure. 
Defining  the 
Alfv\'en velocity associated with the
poloidal magnetic field, 
$v_{Ap}^2\equiv |\nabla \psi|^2/\textstyle \rho$, 
and the Mach number
$
 M^2\equiv v_p^2/v_{Ap}^2 = (F^\prime)^2/\rho
$                                                    
and inserting the expression (\ref{19}) into Eq. (\ref{15}), 
we arrive at the 
{\em elliptic} differential equation
\begin{eqnarray} 
(1-M^2) \Delta^\star \psi - 
	     \frac{1}{2}(M^2)^\prime |\nabla \psi|^2 
					   & &  \jump 
+ \frac{1}{2}\left(\frac{X^2}{1-M^2}\right)^\prime
+ R^2\left(P_s^\prime  -  \Omega \rho^\prime\right) 
+ \frac{R^4}{2}\left(\frac{\rho (\Phi^\prime)^2}{1-M^2}\right)^\prime
    = 0.& & 
						    \label{21a}
\end{eqnarray}
Eq. (\ref{21a}) contains the arbitrary
flux functions $F(\psi)$, $\Phi(\psi)$, $X(\psi)$,
$\rho(\psi)$ and $P_s(\psi)$ which must be found from other
physical considerations. For $\Omega\equiv 0$ Eqs.  (\ref{19}) 
and (\ref{21a}) constitute concise forms of the equilibrium
equations we derived formerly \cite{TaTh} [Eqs. (19) and (22) 
therein]. We note here that the physically plausible  
class of equilibria with isothermal magnetic surfaces, $T=T(\psi)$,
was examined
in our previous studies \cite{ThTa,TaTh} for vanishing gravity. Incompressible 
flows and $T=T(\psi)$ imply that the pressure becomes a  
flux function. For  toroidal
plasmas it was shown that equilibria of this kind are possible.
In addition, steady states with incompressible flows and isobaric magnetic
surfaces were proposed in Ref. [3] as approximate equilibria for the 
Joint-European-Torus tokamak.
It is also noted that for non-ideal plasmas with
arbitrary flows,
i.e. when  a finite resistivity 
is introduced in Eq. (7), 
some of the integrals found in this section
in the form of flux functions are destroyed and the tractability
of an extension of the present investigation becomes questionable.

Recapitulating, the axisymmetric ideal MHD equilibria of a 
gravitating plasma with incompressible flows are governed by the set of 
Eqs. (\ref{2a}) and  
(\ref{21a})
for $\psi$ and $\Omega$, and relation  (\ref{19}) for the pressure. 

Under the transformation  \cite{Cl,ThTa99}
\beq
U(\psi)= \int_0^{\psi}\, [1-M^2(\psi^{\prime})]^{1/2}\, d\psi^\prime,
	 \ \    M^2<1,
						       \label{22}
\eeq
Eq. (\ref{21a})
reduces (after dividing by $(1-M^2)^{1/2}$) to 
\begin{eqnarray} 
 \Delta^\star U  
+ \frac{1}{2}\frac{d}{d U} \left(\frac{X^2}{1-M^2}\right)
+ R^2 \left(\frac{dP_s}{d U}   -  \Omega\frac{d\rho}{d U}\right) 
					   & &  \jump
+ R^4 \frac{d}{d U}
  \left\lbrack\rho \left(\frac{d\Phi}{d U}\right)^2\right\rbrack = 0.
& &                                      
						       \label{23a}
\end{eqnarray}
Eq. (\ref{23a}) is free of the nonlinear term $1/2(M^2)^\prime|\nabla\psi|^2$
and, therefore,  for $M^2<1$  the equilibrium can be determined from the  
more tractable  set of Eqs. (\ref{2a}), (\ref{19}) and (\ref{23a}).   

\begin{center}
{\large\bf 3.\ \ Magnetic-dipole equilibrium equations}
\end{center}                                              

The equilibrium of a  plasma 
confined by the magnetic field
of a current ring lying on the symmetry plane and 
centered at the origin of the system of coordinates
is now considered. 
Employing spherical coordinates $r$, $\theta$ and
$\phi$ with $\mu=\cos \theta$ and $R=r \sin \theta$, 
we seek separable solutions of Eq. (\ref{23a}) of the 
form
\begin{equation}
U(r,\mu)=U_0 H(\mu)\left(\frac{r_0}{r}\right)^\alpha.
					      \label{25}
\end{equation}
Here, $H$ is an unknown function of $\mu$ alone such that
$H(0)=1$, and $U_0$ and  $r_0$ are normalization constants
specifying a reference flux surface location. 
The parameter $\alpha$  plays 
the role of an eigenvalue of  equations (\ref{2a})
and (\ref{23a}). It equals unity or $-2$ in the vacuum limit to recover 
the dipolar solutions $\psi_{vac}\propto (1-\mu^2)/r$ and
$\psi_{vac}\propto (1-\mu^2)r^2$ describing, respectively, the flux
surfaces far away from and close to the origin.
We are interested in configurations symmetric with respect
to the symmetry plane. Accordingly,
the boundary conditions 
\beq
H(\mu\rightarrow 1)\propto 1-\mu,   
\ \ \                                              
\left. \frac{dH}{d\mu}\right|_{\mu =0}=0
					      \label{25a}
\eeq
are chosen to keep the magnetic 
field finite at $\theta=0 $
and parallel to the axis of symmetry at $\theta=\pi/2$,
respectively. 
Using  Eqs. (\ref{7}),  (\ref{10}), (\ref{11}), 
(\ref{22}), and
(\ref{25a}) 
 we find for 
the magnetic field associated with Eq. (\ref{25})   
\barr
{\bf B}& = & \frac{1}{r(1-\mu^2)^{1/2}}
	 \left\lbrack \frac{X}{1-M^2} 
 -r^2(1-\mu^2)\frac{d F}{d U}\frac{d \Phi}{dU}
 \right\rbrack {\bf e}_\phi
						  \jump
& &      + B_0\left(\frac{r_0}{r}\right)^{2+ \alpha}
	   (1-M^2)^{-1/2} \left\lbrack 
	   (1-\mu^2)^{-1/2} H(\mu){\bf e}_\theta  
	+  \frac{1}{\alpha}\frac{d H(\mu)}{d \mu}{\bf e}_r
	   \right\rbrack,
					   \label{25b} 
\earr
where $B_0=\alpha U_0/r_0^2$.   Note  the finite toroidal magnetic 
field induced  by the flow and the poloidal currents as can also be seen  
from Eq. (\ref{11}). 

We now consider a plasma subject  only to gravitating forces 
from a star or
black hole of mass $M_s$ 
placed at $r=0$; the plasma self gravity is neglected. 
Consequently, Eq. (\ref{2a}) decouples from (\ref{23a}) and has the  
solution 
\begin{equation}
\Omega=-\frac{G_0 M_s}{4\pi r}.             
						\label{25c}
\end{equation}
Inspection of Eq. (\ref{23a}) with the gravitation potential
(\ref{25c}) implies that the separable solution
(\ref{25}) is only possible provided
\beq
P_s=P_{s0} \left(\frac{U}{U_0}\right)^{2+4/\alpha},
						\label{25d}
\eeq
\beq
\frac{X^2}{1-M^2} =X_0^2 \left(\frac{U}{U_0}\right)^{2+2/\alpha},
						\label{25d1}
\eeq

\beq
\rho=\rho_0 \left(\frac{U}{U_0}\right)^{2+3/ \alpha},       
						\label{25e}
\eeq
and
\beq
     \rho\left(\frac{d\Phi}{d U}\right)^2 =
      \rho_0\left(\frac{\Phi_0}{U_0}\right)^2
      \left(\frac{U}{U_0}\right)^{2 + 6/\alpha},  
						\label{25f}
\eeq
where $P_{s0}$, $X_0$, $\rho_0$, and $\Phi_0$ are normalization constants 
associated with the reference flux surface $U_0$.
Inserting Eqs. (\ref{25}) and (\ref{25c})-(\ref{25f}) into
Eq. (\ref{23a}), we obtain 
\barr
 \frac{d^2 H}{d \mu^2} + \frac{\alpha(\alpha+1)}{1-\mu^2}H
&=& -\beta_0\alpha(2+\alpha)H^{1+4/\alpha}   
    -\gamma_0\alpha(1+\alpha)(1-\mu^2)^{-1}H^{1+2/\alpha} \jump 
& & - \delta_0\alpha(3+\alpha)(1-\mu^2)H^{1+6/\alpha}  
    - \epsilon_0\alpha(3/2+\alpha)H^{1+3/\alpha}.    \nonu
						  \label{28}
\earr
Here, $\beta_0$, $\gamma_0$, $\delta_0$,  and $\epsilon_0$ are related, 
respectively,
to the static thermal,  poloidal-current, flow and gravitating 
energies normalized to  the poloidal-magnetic-field energy on the reference 
surface:
\beq
\beta_0\equiv\frac{P_{s0}}{B_0^2/2},\ \ 
\gamma_0\equiv \frac{(X_0/r_0)^2/2}{B_0^2/2},  \ \ 
\delta_0\equiv 
\frac{\rho_0(\Phi_0 r_0/U_0)^2/2}{B_0^2/2},  \ \
\epsilon_0\equiv\frac{\rho_0 G_0 M_s/(4\pi r_0)}{B_0^2/2}.
\eeq
Solutions of Eq. (\ref{28})  will be constructed in the
low-energy regime, 
$\beta_0\approx\gamma_0\approx\delta_0\approx \epsilon_0\ll 1$, 
and in the high-energy regime,  $\beta_0\approx\gamma_0\approx\delta_0\approx\epsilon_0\gg 1$. 

\begin{center}
{\large\bf 4.\ \ Solution in the low-energy regime 
($\beta_0\approx\gamma_0\approx\delta_0\approx \epsilon_0\ll 1$)}
\end{center}                                              

For this case  it is convenient to put Eq. (\ref{28})
in the form
\begin{eqnarray}
 \lefteqn{\frac{d}{d \mu}\left\lbrack (1-\mu^2)^2\frac{d}{d\mu}
\left(\frac{H}{1-\mu^2}\right)\right\rbrack
-(1-\alpha)(2+\alpha)H = 
-\beta_0\alpha(2+\alpha)(1-\mu^2)H^{1+4/\alpha}} & & \jump
& &-\gamma_0\alpha(1+\alpha)H^{1+2/\alpha} 
- \delta_0\alpha(3+\alpha)(1-\mu^2)^2 H^{1+6/\alpha}
- \epsilon_0(\frac{3}{2} + \alpha)(1-\mu^2)H^{1+3/\alpha} \nonu
						\label{29}
\end{eqnarray}
where $H \rightarrow 1-\mu^2$ as 
$\beta_0\approx\gamma_0\approx\delta_0\approx \epsilon_0\rightarrow 0$. 
With the use of the boundary conditions (\ref{25a}),
integration of Eq. (\ref{29})
from $\mu=0$ to $\mu=1$ yields
\beq
(2+\alpha)\left\lbrack (1-\alpha)P_1 -\alpha \beta_0 P_2\right\rbrack
-\alpha\left\lbrack \gamma_0(1+\alpha)P_3 + \delta_0 (3+\alpha) P_4 
+ \epsilon_0\left(\frac{3}{2}+ \alpha\right)P_5\right\rbrack = 0
						\label{30}   
\eeq
with
\begin{eqnarray}
P_1=\int_0^1 H d\mu,\  & & P_2=\int_0^1 (1-\mu^2)H^{1+4/\alpha},\ \ 
P_3=\int_0^1 H^{1+2/\alpha}
d\mu  \jump
& &   P_4=\int_0^1 (1-\mu^2)^2H^{1+6/\alpha}\, d\mu, \ \  
      P_5=\int_0^1 (1-\mu^2)H^{1+3/\alpha}\, d\mu.  \nonu   
					       \label{31} 
\end{eqnarray}
To appreciate the impact of finite pressure, finite poloidal current,
flow, and gravity on
the vacuum equilibrium,  the   relations 
$H=1-\mu^2$ and $\alpha\rightarrow 1$
are employed into Eq. (\ref{30}) except for the term $1-\alpha$. 
The departure of $\alpha$ from
unity is then found  
\begin{equation}
1-\alpha=   \frac{512}{1001}\beta_0 + \frac{16}{35}\gamma_0
          + \frac{131072}{230945}\delta_0
	  + \frac{320}{693}\epsilon_0.   
					       \label{32}
\end{equation}

\noindent
Therefore, the modifications of the vacuum equilibrium 
from the finite pressure, poloidal current, flow, and gravity 
are of the  order 
of magnitude of $\beta_0\approx \gamma_0\approx \delta_0\approx\epsilon_0$.
As a result, an analytic solution of Eq. (\ref{29}) can be 
derived by using the vacuum solution $H=1-\mu^2$ in the five terms
in which $H$ appears undifferentiated.  Using the boundary
condition (\ref{25a}) at $\mu=1$ and introducing $t=1-\mu^2=\sin^2\theta$,
integration of Eq. (\ref{29}) from $1$ to $\mu$ yields
\begin{eqnarray}
\frac{d}{d t}\left(\frac{H}{t}\right)&=& 
\frac{1}{4t^2(1-t)^{1/2}}
\left\lbrack 3(1-\alpha)\int_0^t \frac{x\,dx}{(1-x)^{1/2}}
-3\beta_0\int_0^t \frac{x^6\,dx}{(1-x)^{1/2}}  \right.   \jump
& & \left. -2\gamma_0 \int_0^t \frac{x^3\,dx}{(1-x)^{1/2}} 
- 4\delta_0 \int_0^t \frac{x^9\,dx}{(1-x)^{1/2}}
-\frac{5}{2} \epsilon_0 \int_0^t \frac{x^5\,dx}{(1-x)^{1/2}} \right\rbrack.
\nonu
						  \label{33}
\end{eqnarray}
Evaluating the integrals in Eq. (\ref{33}) and 
integrating again, using $H(\mu=0)=1$, we obtain the following 
low-energy solution
valid at all distances from a point dipole:
\barr
\frac{H}{1-\mu^2}&=&1-\left\lbrack \frac{192}{1001}(1-t) 
	     + \frac{160}{2002}(1-t^2) \right. \jump 
 & &          \left. + \frac{20}{429}(1-t^3) + \frac{18}{572}(1-t^4)  
	     + \frac{3}{130}(1-t^5) \right\rbrack
	       \beta_0 \jump                         
 & &         - \left\lbrack\frac{6}{35}(1-t) + 
	       \frac{5}{70}(1-t^2)\right\rbrack   \gamma_0 \jump 
 & &         - \left\lbrack \frac{49152}{230945}(1-t)  
	     + \frac{8192}{92378}(1-t^2) + \frac{7168}{138567}(1-t^3)                
	     + \frac{42256}{923780}(1-t^4)    \right. \jump
 & &  \left. + \frac{2688}{104975}(1-t^5) + \frac{192}{9690}(1-t^6)
	     + \frac{36}{2261}(1-t^7) + \frac{2}{152}(1-t^8)\right\rbrack  
	       \delta_0    \jump
 & &         - \left\lbrack \frac{40}{231}(1-t) +\frac{100}{1386}(1-t^2)
	     + \frac{25}{594}(1-t^3) + \frac{5}{176}(1-t^4)\right\rbrack 
	       \epsilon_0.      
					   \label{34}
\earr
For vanishing poloidal current, flow and gravity  
($\gamma_0=\delta_0=\epsilon_0=0$)
Eq. (\ref{34}) reduces to the
low-pressure static equilibrium solution of Ref. \cite{KrCaHa} 
[Eq. (12) therein].

\begin{center}
{\large \bf 5.\ \ Solution in the high energy regime 
($\beta_0\approx\gamma_0\approx\delta_0\approx \epsilon_0\gg 1$)}
\end{center}                                              

From the low-energy solution we anticipate that
for $\beta_0\approx\gamma_0\approx\delta_0\approx\epsilon_0\gg 1$ it holds that 
$|\alpha|\ll 1$ 
and, consequently, assume that
\beq
\frac{1}{\beta_0}\ll|\alpha|\ll 1,
				      \label{35}
\eeq
to be  verified {\it a posteriori}.
The term $\al(1+\al)H/(1-\mu^2)$ in Eq. (\ref{28})
is small everywhere, 
 via relations
(\ref{25a}) and (\ref{35}), 
and can be neglected.

We first search for solutions in the region 
$0\leq\mu\leq\mu_c\equiv 1/\beta_0\ll 1$.
In this region  the approximations $1-\mu^2\approx 1$,
$H\approx 1$ can be made and, therefore, Eq. (\ref{28}) can be
written in the form
\beq
\frac{d^2 H}{d\mu^2} = -\al\left(2\beta_0 + \gamma_0 + 3\delta_0  
			+ \frac{3}{2}\epsilon_0 \right).
					 \label{36}
\eeq
Integrating twice  Eq. (\ref{36}) with $\left. dH/d\mu\right|_{\mu=0}=0$
and $H(\mu=0)=1$  we obtain the solution
\beq
H=1+\al \left(2\beta_0+\gamma_0 + 3\delta_0+\frac{3}{2}\epsilon_0\right)               
  \mu\left(1-\frac{\mu}{2}\right).  
					     \label{38}
\eeq
Evaluation of Eq. (\ref{38}) at $\mu=\mu_c$  yields
\beq
|\al|= {\cal O}(1-H(\mu_c)) \ll 1,
					    \label{39}
\eeq
consistent with  assumption (\ref{35}).

We now consider the whole regime of variation of $\mu$.
For $0\leq\mu\leq1$ 
the terms on the 
RHS of Eq. (\ref{28}) 
are large at $\mu=0$ 
 and rapidly decrease to zero as $H$ decreases from
$H(\mu=0)=1$ toward $H(\mu=1)=0$ since $|\al|\ll 1$. 
In particular, the variations
of the  terms  $\gamma_0(1-\mu^2)^{-1}\alpha(1+\alpha)H^{1+2/\alpha}$  
and
$\delta_0(1-\mu^2)\alpha(3+\alpha)H^{1+6/\alpha}$
from   $1-\mu^2$ are much weaker than those
from $H^{1+2/\alpha}$ and $H^{1+6/\alpha}$, respectively.
Therefore, in the above  terms,  $1-\mu^2$ can 
be approximated by unity and,
consequently,  Eq. (\ref{28}) can be  written in the form
\beq
\frac{d^2 H}{d\mu^2} = -\al\left(2\beta_0 H^{1+4/\al} + 
                        \gamma_0 H^{1+2/\alpha} + 3\delta_0
		  H^{1+6/\al} + \frac{3}{2}\epsilon_0 H^{1+3/\al}
		  \right).  
						  \label{40}
\eeq
Multiplying Eq. (\ref{40}) by $dH/d\mu$ and integrating from                                         
$\mu=0$ where  $dH/d\mu=0$, we find
\barr
\frac{dH}{d\mu}&=&- |\al| \left\lbrack\beta_0\left(1-H^{2+4/\al}\right)
		  + \gamma_0\left(1-H^{2+2/\al}\right) 
                  + \delta_0\left(1-H^{2+6/\al}\right) \right. \nonu
     & & \left.  + \epsilon_0\left(1-H^{2+3/\al}\right)
		    \right\rbrack^{1/2}   +{\cal O}(\sqrt{|\al|}.
						   \label{41}
\earr
Integration again from $\mu=0$, where $H(\mu=0)=1$, to $\mu$ yields
\barr
|\al|\mu &=& \int_H^1 \left\lbrack \beta_0 \left(1-x^{2+4/\al}\right)
	 + \gamma_0 \left(1-x^{2+2/\al}\right) \right.   \nonu 
  & &   \left.    + \delta_0 \left(1-x^{2+6/\al}\right)   
	 + \epsilon_0 \left(1-x^{2+3/\al}\right) \right\rbrack^{-1/2}\, dx
	   \jump
& & \stackrel{\mu\rar 1}{\longrightarrow}  
    (1-H)(\beta_0+\gamma_0 + \delta_0+ \epsilon_0)^{-1/2}.
						     \label{42}
\earr
To satisfy $H(\mu=1)=0$, Eq. (\ref{42}) requires
\beq
|\al| = (\beta_0 +\gamma_0 + \delta_0 + \epsilon_0)^{-1/2} + 
    {\cal O}(1/(\beta_0 +\gamma_0 +  \delta_0 + \epsilon_0)).
						  \label{43}
\eeq
Note that for $\beta_0\approx\gamma_0\approx\delta_0\approx\epsilon_0\gg 1$,
 $H=1-\mu$ holds  everywhere except in a small region 
$0\leq \mu\leq (\beta_0 + \gamma_0 + \delta_0 + \epsilon_0)^{-1/2}\ll 1$,
where $H$ remains close to unity, but with a large second derivative
of the order of $(\beta_0+\gamma_0 + \delta_0+\epsilon_0)^{1/2}$ 
[see Eq. (\ref{40})].
Eq. (\ref{43}) implies that the distance between adjacent flux 
surfaces at the symmetry plane $\mu=0$ increases as either of $\beta_0$, 
$\gamma_0$,
$\delta_0$, and  $\epsilon_0$ increases. Indeed,
as $\al$ decreases the spacing must adjust to keep $U\approx (r_0/r)^{\al}$
fixed and, therefore, 
the magnetic surfaces become more extended
and localized about the symmetry plane.  
The  resulting  equilibria
 resemble the accretion disks in astrophysics. It is noted that 
Krasheninnikov and Catto \cite{KrCaa,KrCab} came to the conclusion
that in the strong gravity limit ($\epsilon_0\gg 1$),
gravity and  flow affect the flux surfaces but
not the eigenvalue $\alpha$. 
This does not contradict  our conclusion 
because it concerns different density, current and flow regimes, 
i.e. arbitrary  density profiles, vanishing poloidal currents, 
and purely toroidal flows were considered
in Refs. \cite{KrCaa,KrCab} while the present study concerns 
finite  poloidal currents and
magnetic surfaces of constant  density  associated with
incompressible flows with non vanishing toroidal and poloidal
components.
\begin{center} 
{\large \bf 6.\ \ Conclusions}
\end{center}            

It has been shown that the equilibrium of a gravitating axisymmetric 
magnetically confined plasma with incompressible flows is governed by 
a second-order elliptic
differential equation for the poloidal magnetic flux function
$\psi$ [Eq. (\ref{21a})] containing five flux functions
coupled with a Poisson equation for the gravitation potential,
and an algebraic relation for the pressure [Eq. (\ref{19})]. 
The 
above mentioned elliptic equation can be transformed to one  
[Eq. (\ref{23a})] possessing a differential part identical to
that of the Grad-Schl\"uter-Shafranov equation, which permits the 
derivation of  analytic solutions.

Analytic solutions for a plasma confined by a dipolar magnetic field
and subject to  gravitating forces from a massive body have been
obtained in two energy regimes: 
(a) in the low-energy regime
$\beta_0\approx \gamma_0 \approx \delta_0 \approx\epsilon_0\ll 1$, 
where  $\beta_0$, $\gamma_0$, $\delta_0$, and $\epsilon_0$ 
are related to the  
thermal,  poloidal-current, flow and
gravitating  energies normalized to the poloidal-magnetic-field energy, 
respectively,
and (b) in the high-energy regime 
$\beta_0\approx \gamma_0\approx \delta_0 \approx\epsilon_0\gg 1$. 
These solutions   generalize the static magnetic-dipole equilibria 
with vanishing poloidal currents
obtained in Ref. \cite{KrCaHa}. 
It turns out that in the high-energy
regime all four forces, 
 pressure-gradient, toroidal-magnetic-field, inertial and gravitating, 
contribute equally to the
formation of  magnetic surfaces  
very extended and localized about the symmetry plane  such that the
resulting equilibria resemble the accretion 
disks in astrophysics.

Finally, it may be noted that,  in addition to their  
astrophysical concern, the equilibrium investigations of Refs. 
\cite{KrCaHa,KrCaa,KrCab} and of the present work may help in developing 
possible novel magnetic confinement  devices. In this view
further studies on the impact of compressible flows 
or/and self gravity on the equilibrium properties of plasmas confined  
in dipolar magnetic fields are of particular interest.

\begin{center}
 {\large\bf Acknowledgments}
\end{center}

Part of this work was conducted during a visit of one of the authors 
(GNT) to  Max-Planck Institut  f\"ur Plasmaphysik, Garching.
The hospitality of that Institute is greatly appreciated.

\end{document}